# Mobile Botnet Detection: A Deep Learning Approach Using Convolutional Neural Networks

Suleiman Y. Yerima[1] and Mohammed K. Alzaylaee[2]

[1]Cyber Technology Institute
Faculty of Computing, Engineering and Media
De Montfort University, Leicester, United Kingdom
syerima@dmu.ac.uk

[2]Al-Qunfudah College of Computing
Umm Al-Qura University, Saudi Arabia
mkzaylaee@uqu.edu.sa

*Abstract*— Android, being the most widespread mobile operating systems is increasingly becoming a target for malware. Malicious apps designed to turn mobile devices into bots that may form part of a larger botnet have become quite common, thus posing a serious threat. This calls for more effective methods to detect botnets on the Android platform. Hence, in this paper, we present a deep learning approach for Android botnet detection based on Convolutional Neural Networks (CNN). Our proposed botnet detection system is implemented as a CNN-based model that is trained on 342 static app features to distinguish between *botnet apps* and *normal apps*. The trained botnet detection model was evaluated on a set of 6,802 real applications containing 1,929 botnets from the publicly available ISCX botnet dataset. The results show that our CNN-based approach had the highest overall prediction accuracy compared to other popular machine learning classifiers. Furthermore, the performance results observed from our model were better than those reported in previous studies on machine learning based Android botnet detection.

*Keywords—Botnet detection; Deep learning; Convolutional Neural Networks; Machine learning; Android Botnets*

I. INTRODUCTION

Android is now the most widespread mobile operating system worldwide. Over the years the volume of malware targeting Android has continued to grow [1]. This is because it is easier and more profitable for malware authors to target an operating system that is open-source, more prevalent, and does not restrict the installation of apps from any possible source. As a matter of fact, numerous families of malware apps that are capable of infecting Android devices and turning them into malicious bots have been discovered in the wild. These Android bots may become part of a larger botnet that can be used to perform various types of attacks such as Distributed Denial of Service (DDoS) attacks, generation and distribution of Spam, Phishing attacks, click fraud, stealing login credentials or credit card details, etc.

A botnet consists of a number of Internet-connected devices under the control of a malicious user or group of users known as botmaster(s). It also consists of a Command and Control (C&C) infrastructure that enables the bots to receive commands, get updates and send status information to the malicious actors. Since smartphones and other mobile devices are typically used to connect to online services and are rarely switched off, they provide a rich source of candidates for operating botnets. Thus, the term 'mobile botnet' refers to a group of compromised smartphones and other mobile devices that are remotely controlled by botmasters using C&C channels [2], [3].

Nowadays, malicious botnet apps have become a serious threat. Additionally, their increasing use of sophisticated evasive techniques calls for more effective detection approaches. Hence, in this paper we present a deep learning approach that leverages Convolutional Neural Networks (CNN) for Android botnet detection. The CNN model employs 342 static features to classify new or previously unseen apps as either 'botnet' or 'normal'. The features are extracted through automated reverse engineering of the apps, and are used to create feature vectors that feed directly into the CNN model without further pre-processing or feature selection.

We present the design of our CNN-based model for Android botnet detection and evaluate the model on a dataset of real Android apps consisting of 1,929 botnets samples and 4,873 clean samples. Also, we compare the performance of our CNN model to other popular machine learning classifiers including Naïve Bayes, Bayes Net, Decision Tree, Support Vector Machine (SVM), Random Forest, Random Tree, Simple Logistic and Artificial Neural Network (ANN) on the same dataset. The results show that the CNN-based model achieved a botnet detection performance of 98.9% with an F1-score of 0.981, thus outperforming all the other machine learning classifiers. Furthermore, our CNN model shows better performance results compared to other existing studies focusing on Android botnet detection. Some of these studies utilized the same ISCX botnet apps employed in this paper.

The rest of the paper is organized as follows: Section II discusses related works in Android botnet detection; Section III presents the overall system and gives some background on CNN, including a discussion of 1D CNN which is adopted in this study; Section IV presents methodology and the experiments performed; Results of experiments are given in Section V and finally Section VI presents the conclusions of the study and possible future work.

## II. RELATED WORK

In the study conducted by Kadir et al. [4], the objective was to address the gap in understanding mobile botnets and their communication characteristics. Thus, they provided an in-depth analysis of the Command and Control (C&C) and built-in URLs of Android botnets. By combining both static and dynamic analyses with visualization, relationships between the analysed botnet families were uncovered, offering insight into each malicious infrastructure. It is in this study that a dataset of 1929 samples of 14 Android botnet families were compiled and released to the research community. This dataset is known as the ISCX Android botnet dataset and is available from [5]. This paper and several previous works on Android botnets have utilized the full dataset or a subset of it to evaluate proposed Android botnet detection techniques.

Anwar et al. [6] proposed a static approach towards mobile botnet detection where they utilized MD5 hashes, permissions, broadcast receivers, and background services as features. These features were extracted from Android apps to build a machine learning classifier for detecting mobile botnet attacks. They conducted their experiments on 1400 apps from the UNB ISCX botnet dataset together with 1400 benign apps. Their best result was 95.1% classification accuracy with a recall value of 0.827 and a precision value of 0.97.

Paper [7] used machine learning to detect Android botnets based on permissions and their protection levels. The authors initially used 138 features and then added novel features known as protection levels to increase the number of features to 145. Their approach was evaluated on four machine learning algorithms: Random Forest, MLP, Decision Trees and Naïve Bayes. They performed their study on 3270 app instances (1635 benign and 1635 botnets). The botnet apps used were also obtained from the ISCX botnet dataset. The best results came from Random Forest with 97.3% accuracy, 0.987 recall, and 0.958 precision.

In [8] a method was proposed to detect Android botnets based on Convolutional Neural Networks using permissions as features. Applications are represented as images that are constructed based on the co-occurrence of permissions used within the applications. The proposed CNN is a binary classifier that is trained using the images. The authors evaluated their proposed method on 5450 Android applications consisting of 1800 botnet applications from the ISCX dataset. Their results show an accuracy of 97.2% with a recall of 0.96, precision of 0.955 and f-measure of 0.957, which is a promising result considering that only permissions were used in the study.

Paper [9] proposed an Android Botnet Identification System (ABIS) for checking Android applications in order to detect botnets. ABIS utilized both static and dynamic features from API calls, permissions and network traffic. The system is evaluated by using several machine learning algorithms with Random Forest obtaining a precision of 0.972 and a recall of 0.969. In [10], a method is proposed for Android botnet detection based on feature selection and classification algorithms. The paper used 'permissions requested' as features and 'Information gain' to select the most significant permissions. Afterwards, Naïve Bayes, Random Forest and Decision Trees were used to classify the Android apps. Results show Random Forest achieving the highest detection accuracy of 94.6% with the lowest false positive rate of 0.099.

Karim et al [11] proposed DeDroid, a static analysis approach to investigate botnet-specific properties that can be used to detect mobile botnets. They first identified 'critical features' by observing the coding behaviour of a few known malware binaries having C&C features. They then compared these 'critical features' with features of malicious applications from the Drebin dataset [12]. Through this comparison, 35% of the malicious apps in the dataset qualified as botnets. However, closer examination revealed that 90% were confirmed as botnets.

Bernardeschia et al. [13] proposed a method to identify botnets in Android environment through model checking. Model checking is an automated technique for verifying finite state systems. This is accomplished by checking whether a structure representing a system satisfies a temporal logic formula describing their expected behaviour. In [14], Jadhav et al. propose a cloud-based Android botnet detection system which exploits dynamic analysis by using a virtual environment with cluster analysis. The toolchain for the dynamic analysis process within the botnet detection system is composed of strace, netflow, logcat, sysdump, and tcpdump. However, the authors did not provide any experimental results to evaluate the effectiveness of their proposed solution. Moreover, botnets may easily employ different techniques to evade the virtual environment, and code coverage could limit the system's effectiveness [15], [24].

Paper [16] proposed an approach to detect mobile botnets using network features such as TCP/UDP packet size, frame duration, and source/destination IP address. The authors used a set of ML box algorithms and five machine learning classifiers to classify network traffic. The five supervised machine learning approaches include Naïve Bayes, Decision Tree, K-nearest neighbour, Neural Network, and Support Vector Machine. In [17], a method to detect Android botnets based on source code mining and source code metric was proposed. There are also a number of works that have proposed signature based methods for Android botnet detection. These include [18-20]. However, these solutions are likely to suffer from the drawbacks of signature based systems which includes the inability to effectively detect previously unseen botnets.

Unlike most existing studies, our paper proposes a deep learning based Android botnet detection system, using Convolutional Neural Networks. Also, unlike previous studies that utilize only the app permissions, our system is based on 342 features that represent Permissions, API calls, Commands, Extra Files, and Intents. Furthermore, different from the study in [9] which utilized only permissions, we do not convert fea-ture vectors into images prior to model training. Instead our feature vectors are used directly to train 1D CNN models. This makes our approach computationally less demanding.

### III. BACKGROUND

#### A. The CNN-based classification system

The classification system is built by extracting static features from the corpus of botnet and clean samples. To achieve this, we used our bespoke tool built in Python for automated reverse engineering of APKs. With the help of the tool, we extracted 342 features consisting of five different types (see Table 2) from all the training apps. The five feature types include: API calls extracted from the executable; Permissions and Intents from the manifest file; Commands and Extra Files from the APK. These features are represented as vectors of binary numbers with each feature in the vector represented by a '1' or '0'. Each feature vector (corresponding to one application) is labelled with its class. The feature vectors are loaded into the CNN model and used to train the model. After training, an unknown application can be predicted to be either 'clean' or 'botnet' by applying its own extracted feature vector to the trained model. The process is depicted in Figure 1.

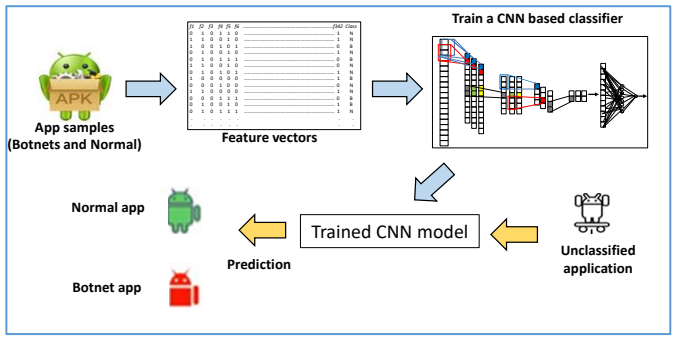

Figure 1: Training and prediction with the CNN-based botnet detection system.

#### B. Convolutional Neural Networks (CNN)

A CNN is a deep learning technique that belongs to the family of Artificial Neural Networks. It works well for identifying simple patterns in the data which will then be used to form more complex patterns in higher layers. Two types of layers are typically used for building CNNs; convolutional layers and pooling layers. The role of the convolutional layer is to detect local conjunctions of features from the previous layer, while the role of the pooling layer is to merge semantically similar features into one [21].

Generally, the convolutional layer extracts the optimal features while the pooling layer reduces the dimensions of those features that it receives from the convolutional layer (or another preceding pooling layer). At the tail end of the model, fully connected (dense) layer(s) are typically used for classification. Depending on the characteristics of the dataset, the performance of the CNN may be influenced by the number of layers, number of filters (kernels) or the size of the filters. Generally, more and more abstract features are extracted in the deeper layers of the CNN, hence, the number of layers required depends on the complexity and non-linearity of the data being analysed. Furthermore, the number of filters in each stage determines the number of features extracted. Computational complexity increases with more layers and higher numbers of filters. Also, with more complex architectures, there is the possibility of training an overfitted model which results in poor prediction accuracy on the testing set(s). To reduce overfitting, techniques such as 'dropout' [22] and 'batch regularization' are implemented during training of our models.

#### C. One Dimensional Convolutional Neural Networks

Although CNN is more commonly applied in a multi-dimensional fashion and has thus found success in image and video analysis-based problems, they can also be applied to one-dimensional data. Datasets that possess a one-dimensional structure can be processed using a one-dimensional convolutional neural network (1D CNN). The key difference between a 1D and a 2D or 3D CNN is the dimensionality of the input data and how the filter (feature detector) slides across the data. For 1D CNN, the filters only slide across the input data in one direction. A 1D CNN is quite effective when you expect to derive interesting features from shorter (fixed-length) segments of the overall feature set, and where the location of the feature within the segment is not of high relevance.

The use of 1D CNN can be commonly found in NLP applications. Similarly, 1D CNN is applicable to datasets containing vectorised data being used to characterize the items to be predicted (e.g. an Android application). The 1D CNN could be used to extract potentially more discriminative feature representations that describe any existing patterns or relationships within segments of the vectors characterizing each entity in the dataset. These new features are then fed into a classifier (e.g. a fully connected neural network layer) which will in turn use the derived features in making a final classification decision. Hence, in this scenario, the convolutional layers can be considered as a feature extractor that eliminates the need for feature ranking and selection. The CNN model developed in this paper is applied to vectorised data characterizing the Android applications, in order to derive a trained model that can detect new Android botnet apps with very high accuracy.

#### D. Key elements of our proposed CNN architecture

Our proposed CNN architecture is a 1D CNN consisting of two convolutional layers and two max pooling layers. These are followed by a fully connected layer of *N* units, which is in turn connected to a final classification layer containing one neuron with a *sigmoid* activation function.

The sigmoid activation function is given by: $S = \frac{1}{1+ e^{-x}}$

The final classification layer generates an outcome corresponding to the two classes i.e. 'botnet' or 'normal'. The convolutional layers utilize the *ReLU* (Rectified Linear Units) activation function given by: $f(x) = \max(0, x)$. *ReLU* helps to mitigate vanishing and exploding gradient issues [23]. It has been found to be more efficient in terms of time and cost for training huge data in comparison to classical non-linear activa-tion functions such as Sigmoid or Tangent functions [24]. A simplified view of our architecture is shown in Figure 2.

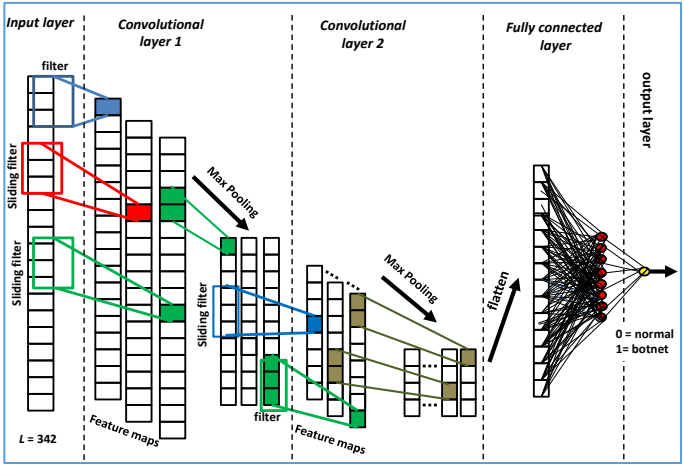

Figure 2: Overview of the implemented 1D CNN model for Android application classification to detect botnets.

## IV. METHODOLOGY AND EXPERIMENTS

In this section we present the experiments undertaken to evaluate the CNN models developed in this paper. Our models were implemented using Python and utilized the Keras library with TensorFlow backend. Other libraries used include Scikit Learn, Seaborn, Pandas, and Numpy. The model was built and evaluated on an Ubuntu Linux 16.04 64-bit Machine with 4GB RAM.

### A. Problem definition

Let $A = \{a_1, a_2, \ldots a_n\}$ be a set of apps where each $a_i$ is represented by a vector containing the values of $n$ features (where $n=342$). Let $a = \{f_1, f_2, f_3 \ldots f_n, cl\}$ where $cl \in \{botnet, normal\}$ is the class label assigned to the app. Thus, $A$ can be used to train the model to learn the behaviours of *botnet* and n*ormal* apps respectively. The goal of a trained model is then to classify a given unlabelled app $A_{unknown} = \{f_1, f_2, f_3 \ldots f_n, ?\}$ by assigning a label $cl$, where $cl \in \{botnet, normal\}$.

### B. Dataset

In this study we used the Android dataset from [5], which is known as the ISCX botnet dataset. The ISCX dataset contains 1,929 botnet apps (from 14 different families) and has been used in previous works including [4], [7-10], and [17]. The botnet families are shown in Table 1. A total of 4,873 clean apps were used for the study in this paper and these were labelled under the category 'normal' to facilitate supervised learning when training the CNN and other machine learning classifiers. The clean apps were obtained from different categories of apps on the Google Play store and verified to be non-malicious by using VirusTotal.

The 342 static features extracted from the apps for model training were of 5 types: (a) API calls (b) commands (c) permissions (d) Intents (e) extra files. The 'API calls' and 'permissions' accounted for most of the features. From Table 2, it can be seen that there were 135 'API calls' related features and 130 'permissions' features, while intents accounted for 53 features. Some of the features are shown in Table 3.

Table 1: Botnet dataset composition.

| Botnet Family | Number of samples |
|---|---|
| Anserverbot | 244 |
| Bmaster | 6 |
| Droiddream | 363 |
| Geinimi | 264 |
| Misosms | 100 |
| Nickyspy | 199 |
| Notcompatible | 76 |
| Pjapps | 244 |
| Pletor | 85 |
| Rootsmart | 28 |
| Sandroid | 44 |
| Tigerbot | 96 |
| Wroba | 100 |
| Zitmo | 80 |
| **Total** | **1929** |

Table 2: The five different types of features used to train the CNN model.

| Feature type | Number |
|---|---|
| API calls | 135 |
| Permissions | 130 |
| Commands | 19 |
| Extra files | 5 |
| Intents | 53 |
| **Total** | **342 features** |

Table 3: Some of the prominent static features extracted from Android applications for training the CNN model to detect Android Botnets.

| Feature name | Type |
|---|---|
| TelephonyManager.*getDeviceId | API |
| TelephonyManager.*getSubscriberId | API |
| abortBroadcast | API |
| SEND_SMS | Permission |
| DELETE_PACKAGES | Permission |
| PHONE_STATE | Permission |
| SMS_RECIVED | Permission |
| Ljava.net.InetSocketAddress | API |
| READ_SMS | Permission |
| Android.intent.action.BOOT_COMPLETED | Intent |
| io.File.*delete( | API |
| chown | Command |
| chmod | Command |
| Mount | Command |
| .apk | Extra File |
| .zip | Extra File |
| .dex | Extra File |
| .jar | Extra file |
| CAMERA | Permission |
| ACCESS_FINE_LOCATION | Permission |
| INSTALL_PACKAGES | Permission |
| android.intent.action.BATTERY_LOW | Intent |
| .so | Extra File |
| android.intent.action.POWER_CONNECTED | Intent |
| System.*LoadLibrary | API |

## C. Experiments to evaluate the proposed CNN based model

In order to investigate the performance of our proposed model, we performed different sets of experiments. Table 4 shows the configuration of the CNN model. The 1D CNN model consists of two pairs of convolutional and maxpooling layers as shown in Figure 2. The output of the second max pooling layer is flattened and passed on to a fully connected layer with 8 units. This is in turn connected to a sigmoid activated output layer containing one unit.

The first set of experiments was aimed at evaluating the impact of number of filters on the model's performance. The second set of experiments was performed to evaluate the effect of varying the length of the filters. In the third, we investigate the impact of the maxpooling size on performance.

Table 4: Summary of model configurations.

| Model design summary -1D CNN |
|---|
| **Input layer:** Dimension = 342   (feature vector size) |
| **1D Convolutional layer:** 4, 8, 16, 32, 64 filters, size = 4, 8, 16, 32, 64 (with number of filters =32) |
| **MaxPooling layer:** Size =2, 4, 8, 16 (with number of filters =32) |
| **1D Convolutional layer:** 4, 8, 16, 32, 64 filters, size = 4, 8, 16, 32, 64 (with number of filters =32) |
| **MaxPooling layer:** Size  =2, 4, 8, 16 (with number of filters =32) |
| **Fully Connected (Dense) layer: 8 units**, activation=ReLU |
| **Output layer: Fully Connected  layer**; 1 unit, activation=sigmoid |

In order to measure model performance, we used the following metrics: *Accuracy, precision, recall and F1-score*. The metrics are defined as follows (taking botnet class as positive):

- *Accuracy*: Defined as the ratio between correctly predicted outcomes and the sum of all predictions. It is given by: $\frac{TP+TN}{TP+TN+FP+FN}$
- *Precision*: All true positives divided by all positive predictions. i.e. Was the model right when it predicted positive? Given by: $\frac{TP}{TP+FP}$
- *Recall*: True positives divided by all actual positives. That is, how many positives did the model identify out of all possible positives? Given by: $\frac{TP}{TP+FN}$
- *F1-score*: This is the weighted average of precision and recall, given by: $\frac{2 \times Recall \times Precision}{Recall+Precision}$

Where TP is true positives; FP is false positives; FN is false negatives, while TN is true negatives (all w.r.t. the botnet class). All the results of the experiments are from 10-fold cross validation where the dataset is divided into 10 equal parts with 10% of the dataset held out for testing, while the models are trained from the remaining 90%.  This is repeated until all of the 10 parts have been used for testing. The average of all 10 results is then taken to produce the final result. Also, during the training of the CNN models (for each fold), 10% of the training set was used for validation.

## V.     RESULTS AND DISCUSSIONS

### A. Varying the numbers of filters.

In this section, we examine the results from experimenting with different numbers of filters. In our model, we kept the number of filters in both convolutional layers the same.  Table 5 shows the results from running the 1D CNN model with different numbers of filters. From the table, it is evident that the number of filters had an effect on the performance of the model. When increased from 4 to 8, there is an improvement in performance. The performance does not improve until we reach 32 filters. It then drops again when we increase this to 64. Based on these results we select 32 filters as the optimal configuration parameter for the model's number of filters. Notice the increase in the number of training parameters as the number of filters is increased, and for 32 filters, the training of 25,625 parameters is required. With 32 filters we obtain a classification accuracy of 98.9% compared to 98.6% that is obtained with 4 filters. Nevertheless, the results obtain with 4 filters were still acceptable.

*1) Training epochs, loss and accuracy graphs.*

Figures 3 and 4 shows the typical outputs obtained with the validation and training sets during the training epochs. From Fig. 3, it can be seen that the validation loss is generally fluctuating from one training epoch to another after an initial drop. During each epoch, a model is trained and the validation loss and accuracy are recorded. Our goal is to obtain the model with the least validation loss because we assume this will be the 'best' model that fits the training data. Thus, at every epoch, the validation loss is compared to previous ones and if the current one is lower, the corresponding model is saved as the best model. We implemented a 'stopping criterion' which will stop the training once no improvement in performance is observed within 100 epochs. For example in Figure 3, the best model was obtained with the least validation loss of 0.00531 at epoch 45. For the next 100 epochs validation loss did not improve, hence the training was stopped. Figure 4 shows the corresponding accuracy behaviour observed from epoch to epoch.

Table 5: Number of filters vs. model performance. Length of filters used= 4 for first layer and =4 for second layer; dense layer = 8 units; validation split=10%.

| Number of Filters | 4 | 8 | 16 | 32 | 64 |
|---|---|---|---|---|---|
| Accuracy | 0.986 | 0.988 | 0.988 | **0.989** | 0.987 |
| Precision | 0.978 | 0.980 | 0.980 | **0.983** | 0.980 |
| Recall | 0.974 | 0.977 | 0.976 | **0.978** | 0.975 |
| F1-score | 0.976 | 0.978 | 0.978 | **0.981** | 0.977 |
| Num. training parameters | 2777 | 5,657 | 11,801 | 25,625 | 59,417 |

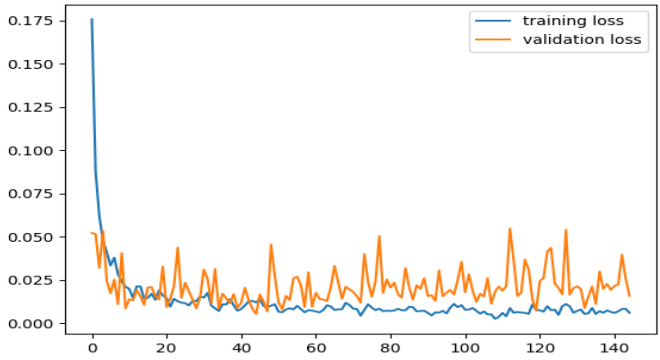

Figure 3: Training and validation losses at different epochs up to 145. A stopping criterion of 100 is used to obtain the model with the least validation loss.

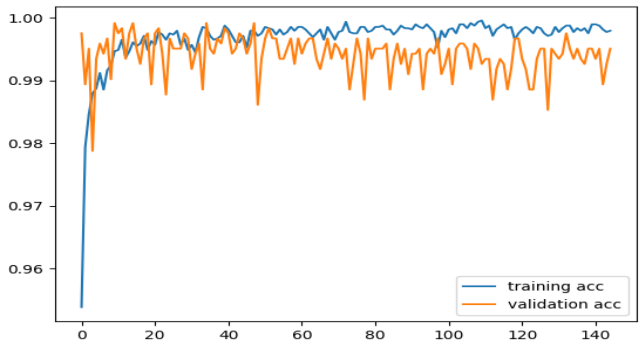

Figure 4: Training and validation accuracies at different epochs up to 145. These plots correspond to the training and validation losses depicted in Figure 3.

### B. Varying the length of the filters.

In this section we examine the effect of the length of filters on the performance of the model while the number of filters is fixed at 32 in each convolutional layer. The length is varied from 4, 8, 16, 32, to 64 respectively (as shown in Table 6). The number of units in the dense layer was fixed at 8. The results indicate that the length of the filters does not appear to have much of an impact on the overall classification accuracy and F1-score performance, when increased. However, the least filter length of 4 achieves the highest accuracy and F1-score. Note that as we increase the length of the filters, the number of parameters to be trained increases (from 25,652 for length=4 to 77,465 for length=64).

The lack of improvement with the length of filters may be attributed to larger number of parameters leading to overfitting the model to the training data thereby reducing its generalization capability. This in turn leads to degraded performance when tested on new data. Basically, what these results show is that when the training parameters increase beyond a certain limit, the model becomes too complex for the data and this leads to overfitting. This becomes evident in lack of improvement or degradation in performance when tested on previously unseen data.

Table 6: Length of filters vs. model performance. Number of filters used= 32 in both first and second convolutional layers; dense layer = 8 units; validation split=10%.

| Length of filters | 4 | 8 | 16 | 32 | 64 |
|---|---|---|---|---|---|
| Accuracy | 0.989 | 0.988 | 0.988 | 0.988 | 0.988 |
| Precision | 0.983 | 0.979 | 0.980 | 0.981 | 0.983 |
| Recall | 0.978 | 0.977 | 0.978 | 0.979 | 0.974 |
| F1-score | 0.981 | 0.978 | 0.979 | 0.979 | 0.978 |
| Training parameters | 25,625 | 29,081 | 35,993 | 49,817 | 77465 |

### C. Varying the Maxpooling parameter

The results of the third set of experiments are discussed here. The goal is to investigate the effect of changing the maxpooling parameter. This corresponds to a subsampling ratio of 2, 4, 6, and 8 respectively as shown in Table 7. A value of 2 means the next layer will be half the dimension of the previous one, etc. Note that the maxpooling layer can be considered a feature reduction layer that also helps to alleviate overfitting since it progressively reduces the number of parameters that need to be trained. The other parameters were fixed as follows: Number of filters in both convolutional layers = 32; Length of convolutional filters = 4; number of units in dense layer=8.

It can be seen from Table 7 that as we increase the maxpooling parameter, the total number of training parameters is reduced. At the same time, we witness a progressive decline in overall performance. Therefore, for our CNN model designed to classify applications into 'botnet' and 'normal', the optimal subsampling ratio for both layers is 2.

Table 7: Maxpooling parameter vs. model performance. Length of filters used=4 for both convolutional layers; number of filters =32 for both layers; dense layer = 8 units; validation split=10%.

| Maxpooling parameter/Subsampling ratio | 2 | 4 | 6 | 8 |
|---|---|---|---|---|
| Accuracy | **0.989** | 0.987 | 0.983 | 0.978 |
| Precision | **0.983** | 0.982 | 0.974 | 0.971 |
| Recall | **0.978** | 0.973 | 0.967 | 0.948 |
| F1-score | **0.981** | 0.978 | 0.970 | 0.959 |
| Training Parameters | 25,625 | 9497 | 6,425 | 5,401 |

### D. CNN performance vs. other machine learning classifiers: 10 fold cross validation results.

In Table 8, the performance of the CNN model developed in this paper is compared to other machine learning classifiers: Naïve Bayes, SVM, Random Forest, Artificial Neural Network, J48, Random Tree, REPtree, and Bayes Net. Figure 5 shows the F1-scores of the classifiers, where CNN has the highest F1-score (0.981), followed by SVM (0.976), SL (0.973), ANN (0.973) and Random Forest (0.973). Bayes Net had the least F1-score of 0.781. Table 8 shows that the recall of CNN is 0.978 which indicates that it has the best botnet detection performance than the other classifiers. Note that the ANN was a back propagation neural network built with a single hidden layer consisting 32 units (neurons). The sigmoid activation function was used within the neurons. This ANN represented the application of a neural network without deep learning. The ANN showed no significant improvement in the results when the number of units in the hidden layer was increased beyond 32.

Table 8: Comparison of our CNN results with results from other ML classifiers.

|  | ACC | Prec. | Rec. | F1 |
|---|---|---|---|---|
| **Naïve Bayes** | 0.872 | 0.728 | 0.874 | 0.795 |
| **SVM** | 0.987 | 0.980 | 0.973 | 0.976 |
| **RF** | 0.985 | 0.982 | 0.965 | 0.973 |
| **ANN** | 0.985 | 0.982 | 0.965 | 0.973 |
| **SL** | 0.984 | 0.983 | 0.963 | 0.973 |
| **J48** | 0.981 | 0.974 | 0.958 | 0.966 |
| **Random Tree** | 0.972 | 0.948 | 0.955 | 0.951 |
| **REPTree** | 0.979 | 0.973 | 0.954 | 0.963 |
| **Bayes Net** | 0.867 | 0.736 | 0.832 | 0.781 |
| **CNN** | **0.989** | **0.983** | **0.978** | **0.981** |

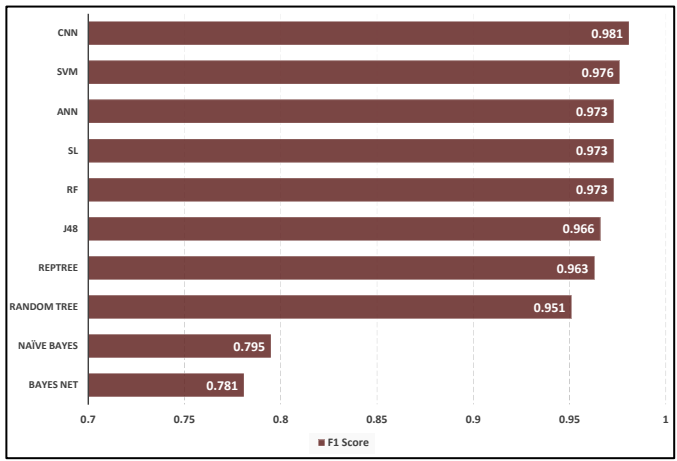

Figure 5: F1-score of CNN vs other ML classifiers.

*E. Comparison with other works on Android botnet detection.*

In Table 9, we present a comparison of our results with those reported in other papers that focus on Android botnet detection. Note that all the papers mentioned in the table have used the ISCX botnet dataset for their work. In our study we utilized the entire 1929 samples within the dataset. In the second column of the table, the numbers of botnet samples and benign samples used in the papers are shown, while the other columns contain the performance results. Not all of the performance metrics we have used are reported in every paper. Nevertheless, it is clear that our CNN model obtained better overall accuracy, F1 and recall than the other works.

Table 9: performance comparisons with other works. Note that all of the papers used botnets samples from the ISCX dataset.

| Paper reference | Botnets /Benign | ACC (%) | Rec. | Prec. | F1 |
|---|---|---|---|---|---|
| Hojjatinia et al. [8] | 1800/3650 | 97.2 | 0.96 | 0.955 | 0.957 |
| Tansettanakorn et al. [9] | 1926/150 | - | 0.969 | 0.972 | - |
| Anwar et. al [6] | 1400/1400 | 95.1 | 0.827 | 0.97 | - |
| Abdullah et al. [10] | 1505/850 | - | 0.946 | 0.931 | - |
| Alqatawna & Faris [7] | 1635/1635 | 97.3 | 0.957 | 0.987 | - |
| **This paper** | **1929/4873** | **98.9** | **0.978** | **0.983** | **0.981** |

VI. CONCLUSIONS AND FUTURE WORK

In this paper, we proposed a deep learning model based on 1D CNN for the detection of Android botnets. We evaluated the model through extensive experiments with 1,929 botnet apps and 4,387 clean apps. The model outperforms several popular machine learning classifiers evaluated on the same dataset. The results (Accuracy: 98.9%; Precision: 0.983; Recall: 0.978; F1-score: 0.981) indicate that our proposed CNN based model can be used to detect new, previously unseen Android botnets more accurately than the other models. For future work, we will aim to improve the model training process by automating the search and selection of the key influencing parameters (i.e. number of filters, filter length, and number of fully connected (dense) layers) that jointly result in the optimal performing CNN model.